\documentclass[twocolumn,twoside,slac_two]{revtex4}
\usepackage{graphicx}
\usepackage{fancyhdr}
\usepackage{hyperref}
\hypersetup{
    colorlinks=true,
    urlcolor=magenta,
}
\newcommand{\cl}[1]{\centerline{#1}}
\begin{document}

\title{First results of the Lomonosov TUS and GRB experiments}

\author{%
	S.V.~Biktemerova$^{b}$,
	A.V.~Bogomolov$^{a}$,
	V.V.~Bogomolov$^{a}$,
	A.A.~Botvinko$^{c}$,
	A.J.~Castro-Tirado$^{d}$,
	E.S.~Gorbovskoy$^{a}$,
	N.P.~Chirskaya$^{a}$,
	V.E.~Eremeev$^{a}$,
	G.K.~Garipov$^{a}$,
	V.M.~Grebenyuk$^{b,e}$,
	A.A.~Grinyuk$^{a}$,
	A.F.~Iyudin$^{a}$,
	S.~Jeong$^{f}$,
	H.M.~Jeong$^{f}$,
	N.L.~Jioeva$^{a}$,
	P.S.~Kazarjan$^{a}$,
	N.N.~Kalmykov$^{a}$,
	M.A.~Kaznacheeva$^{a}$,
	B.A.~Khrenov$^{a}$,
	M.B.~Kim$^{f}$,
	P.A.~Klimov$^{a}$,
	E.A.~Kuznetsova$^{a}$,
	M.V.~Lavrova$^{a}$,
	J.~Lee$^{f}$,
	V.M.~Lipunov$^{a}$,
	O.~Martinez$^{g}$,
	I.N.~Mjagkova$^{a}$,
	M.I.~Panasyuk$^{a}$,
	I.H.~Park$^{f}$,
	V.L.~Petrov$^{a}$,
	E.~Ponce$^{g}$,
	A.E.~Puchkov$^{c}$,
	H.~Salazar$^{g}$,
	O.A.~Saprykin$^{c}$,
	A.N.~Senkovsky$^{c}$,
	S.A.~Sharakin$^{a}$,
	A.V.~Shirokov$^{a}$,
	S.I.~Svertilov$^{a}$,
	A.V.~Tkachenko$^{b}$,
	L.G.~Tkachev$^{b,e}$,
	I.V.~Yashin$^{a}$,
	M.Yu.~Zotov$^a$
}

\affiliation{$^a$Lomonosov Moscow State University, GSP-1, Leninskie
	Gory, Moscow, 119991, Russia}
\affiliation{$^b$Joint Institute for Nuclear Research, Joliot-Curie, 6,
	Dubna, Moscow region, Russia, 141980}
\affiliation{$^c$Space Regatta Consortium, ul. Lenina, 4a,
	141070 Korolev, Moscow region, Russia}
\affiliation{$^d$Instituto de Astrofisica de Andalucia (IAA-CSIC), P.O.Box 03004,
	E-18080 Granada, Spain}
\affiliation{$^e$Dubna State University, University str., 19, Bld.1,
	Dubna, Moscow region, Russia}
\affiliation{$^f$Department of Physics and ISTS, Sungkyunkwan
	University,	Seobu-ro 2066, Suwon, 440-746 Korea}
\affiliation{$^g$Benem\'{e}rita Universidad Aut\'{o}noma de Puebla,
	4 sur 104 Centro Hist\'orico C.P. 72000, Puebla, Mexico}

\begin{abstract}

	On April 28, 2016, the Lomonosov satellite, equipped with a number of
	scientific instruments, was launched into orbit.  Here we present
	briefly some of the results obtained with the first orbital telescope of
	extreme energy cosmic rays TUS and by a group of detectors aimed at
	multi-messenger observations of gamma-ray bursts.

\end{abstract}

\maketitle
\thispagestyle{fancy}

\section{Introduction}

``Lomonosov'' is an orbital scientific laboratory developed at
Lomonosov Moscow State University in close collaboration with
Joint Institute for Nuclear Research (Russia),
University of California, Los Angeles (USA),
Sungkyunkwan University (Republic of Korea),
Benem\'erita Universidad Aut\'onoma de Puebla (Mexico), 
Russian space industry organizations and a few other partners.
The primary objectives of the Lomonosov mission include registration of:

\begin{itemize}\parskip=0pt

	\item extreme energy cosmic rays (EECRs, energies above
		$\sim50$~EeV),

	\item gamma-ray bursts (GRBs) in visible, UV, gamma-, and X-rays,

	\item transient luminous events (TLEs) in the upper atmosphere,

	\item energetic trapped and precipitated  radiation (electrons and
		protons) at a low-Earth orbit in connection with global
		geomagnetic disturbances.

\end{itemize}

To reach the goals, Lomonosov is equipped with a whole number of
different instruments~\cite{Lomonosov2013}.
It was put into orbit on April 28, 2016, from the new Russian space
launching site Vostochny.  The
satellite has a sun-synchronous orbit with an inclination of
$97^\circ\!\!.3$, a period of $\approx94$~min, and a height of about
500~km.

In what follows, we briefly present some results obtained with the
first orbital detector of EECRs, TUS, and a group of instruments aimed
at detecting GRBs.

\section{Results from TUS}

TUS (Tracking Ultraviolet Set-up) is the first
detector developed for registering extreme energy cosmic rays from
space~\cite{2001AIPC..566...57K,TUS-ecrs2012}.
It consists of two main parts: a
parabolic mirror-concentrator of the Fresnel type and a square-shaped
256-pixel photodetector in the focal plane of the mirror. The mirror has
an area of about 2~m$^2$. The field of view (FOV) equals approximately
$80~\mathrm{km}\times80~\mathrm{km}$ at sea level with a spatial
resolution of 5~km.
Pixels of the TUS photodetector are
photomultiplier tubes Hamamatsu R1463 with multialkali cathode of 13~mm
diameter. The pixel wavelength band 240--400~nm is limited by a UV filter
and by PMT quantum efficiency. Light guides with square entrance
apertures (15~mm$\times$15~mm) and circular outputs are employed to
fill uniformly the detector's FOV.

The TUS electronics can operate in four modes intended for detecting
various fast UV phenomena in the atmosphere at different time scales
with different time sampling. The main mode is aimed at registering
extensive air showers (EASs) born by EECRs and has a time sampling of
0.8~$\mu$s. Two other modes have time sampling of 25.6~$\mu$s and 0.4~ms
for studying TLEs of different kinds slower than elves: sprites, blue
jets, gigantic jets, etc.  The fourth mode has a time sampling of 6.6~ms
for detecting micro-meteors, space debris and thunderstorm activity at a
longer time scale. In each mode, waveforms consist of 256 time samples.
The trigger algorithm consists of two levels. The first level trigger
decision is based on a comparison of fixed length sums of
analog-to-digital converter (ADC) counts calculated for each pixel with
a threshold level that depends on a similar value obtained for the
background noise. At the second level trigger, the geometry and number
of hit pixels are analyzed.
Accuracy of the trigger time stamps equals 1~s.
TUS operates during nocturnal segments of
the orbit.

All four modes were tested in the first months of operation. The
main observational time was dedicated to the EAS mode. The
measurements demonstrated an unexpectedly rich variety of UV radiation
in the atmosphere, including anthropogenic and auroral lights,
distant thunderstorm effects, etc., which considerably limit
time for EAS observations. In what follows, we present a few examples of
different phenomena registered with TUS.

More than 80\% of all TUS records consist of noise-like events that are
likely to be caused by random fluctuations of the background.  It is
sometimes possible to identify events of an anthropogenic origin among
them---light of cities, industrial sites, airports, etc. These events
have strongly non-uniform illumination of the focal plane.  A frequency
modulation of the signal at 100~Hz or 120~Hz is observed for such events
in case they are registered with a temporal resolution of 0.4~ms.

\begin{figure}[!ht]
	\cl{\includegraphics[width=80mm]{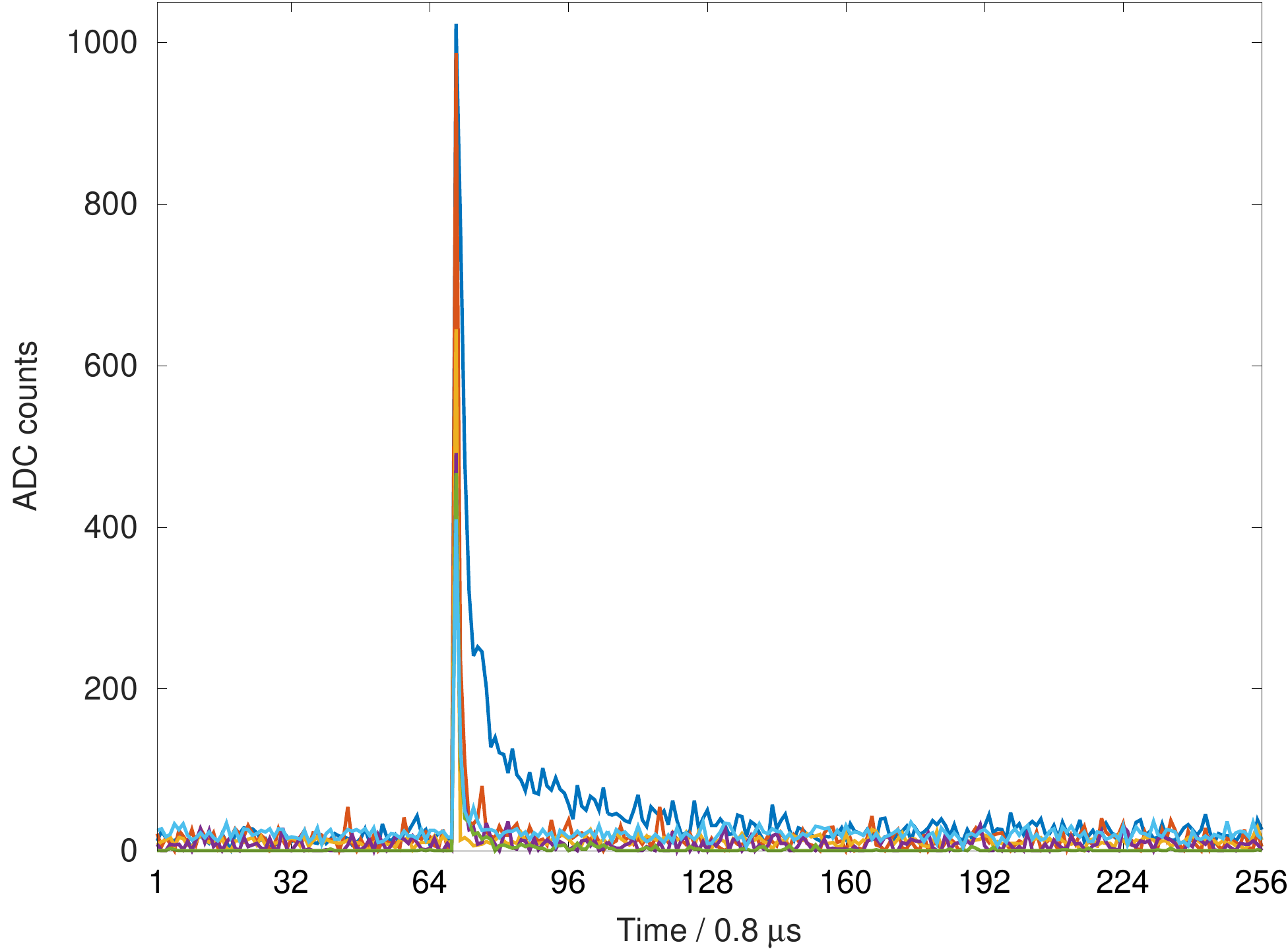}}
	\cl{\includegraphics[width=0.4\textwidth]{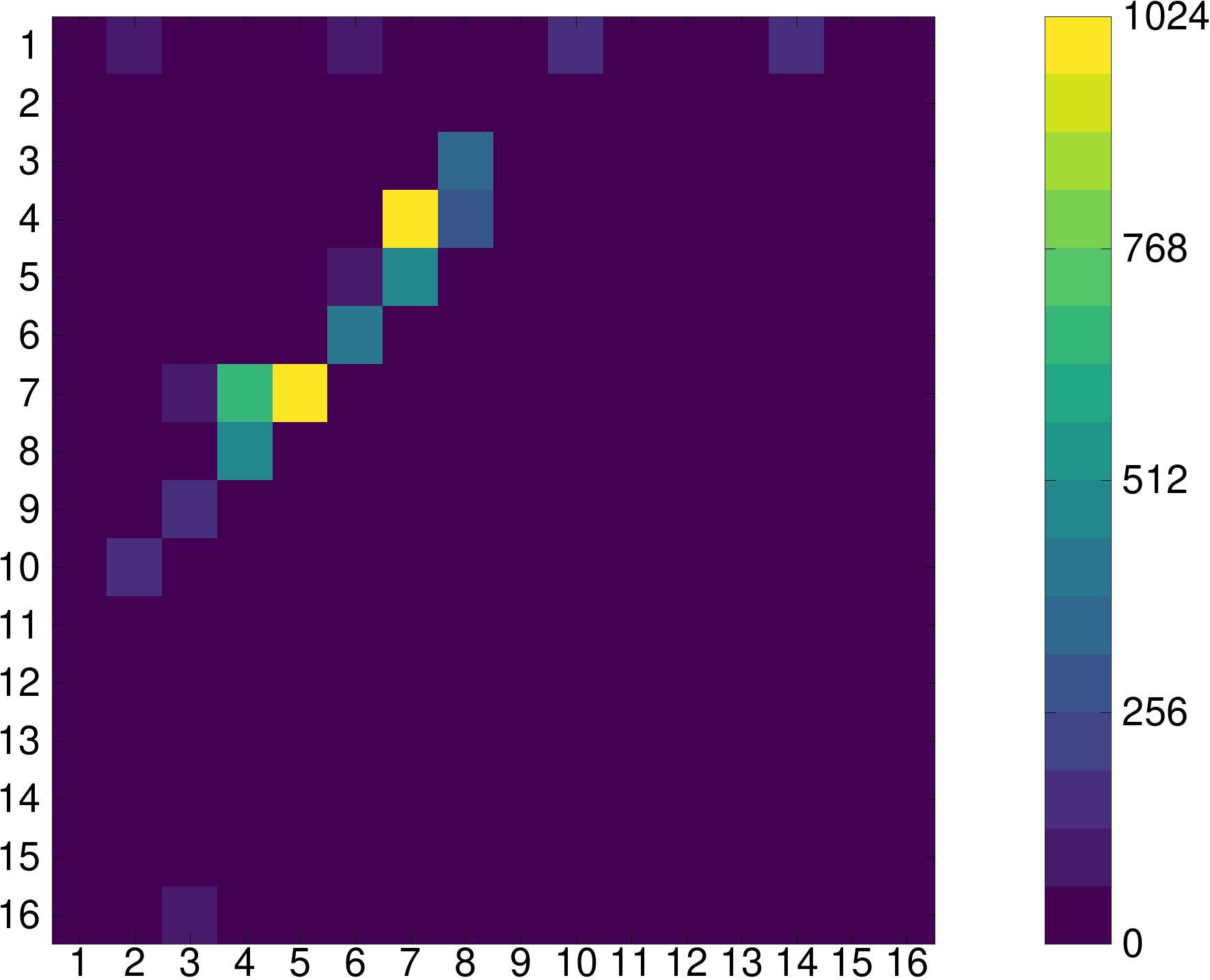}}
	\caption{Example of a track-like event.
	Top: waveforms of six PMTs with the biggest ADC
	counts. Colours denote different pixels. Bottom: snapshot of the focal
	surface at the moment of the peak. Colours denote real ADC counts.}
	\label{fig:track}
\end{figure}

Another major group of events ($\sim14$\%) represents instant (i.e.,
happening in one or, rarely, two time samples of 0.8~$\mu$s) and usually
intensive flashes that produce tracks or, sometimes, small spots in the
focal surface. An example is shown in Fig.~\ref{fig:track}.  Preliminary
simulations~\cite{tracks-izvran} performed with the Geant4
framework~\cite{Geant4} and the geographical distribution of the events
support the hypothesis that they are due to charged particles passing
through the UV glass filters and PMTs of the focal plane.  A detailed
analysis is in progress.

Another group of events registered in the EAS mode consists of records
with monotonously increasing ADC counts with comparatively slow rise
time (100~$\mu$s).  Such flashes typically evolve simultaneously in the
majority of pixels presenting an almost uniform illumination of the
focal plane. An analysis of the geographical distribution of the flashes
and their comparison with data from the World-Wide Lightning Location
Network (\href{http://www.wwlln.net/}{WWLLN}) provide evidence for their
close relation to simultaneous thunderstorm activity, possibly far from the
FOV of TUS.

\begin{figure}[!ht]
	\cl{\includegraphics[width=0.4\textwidth]{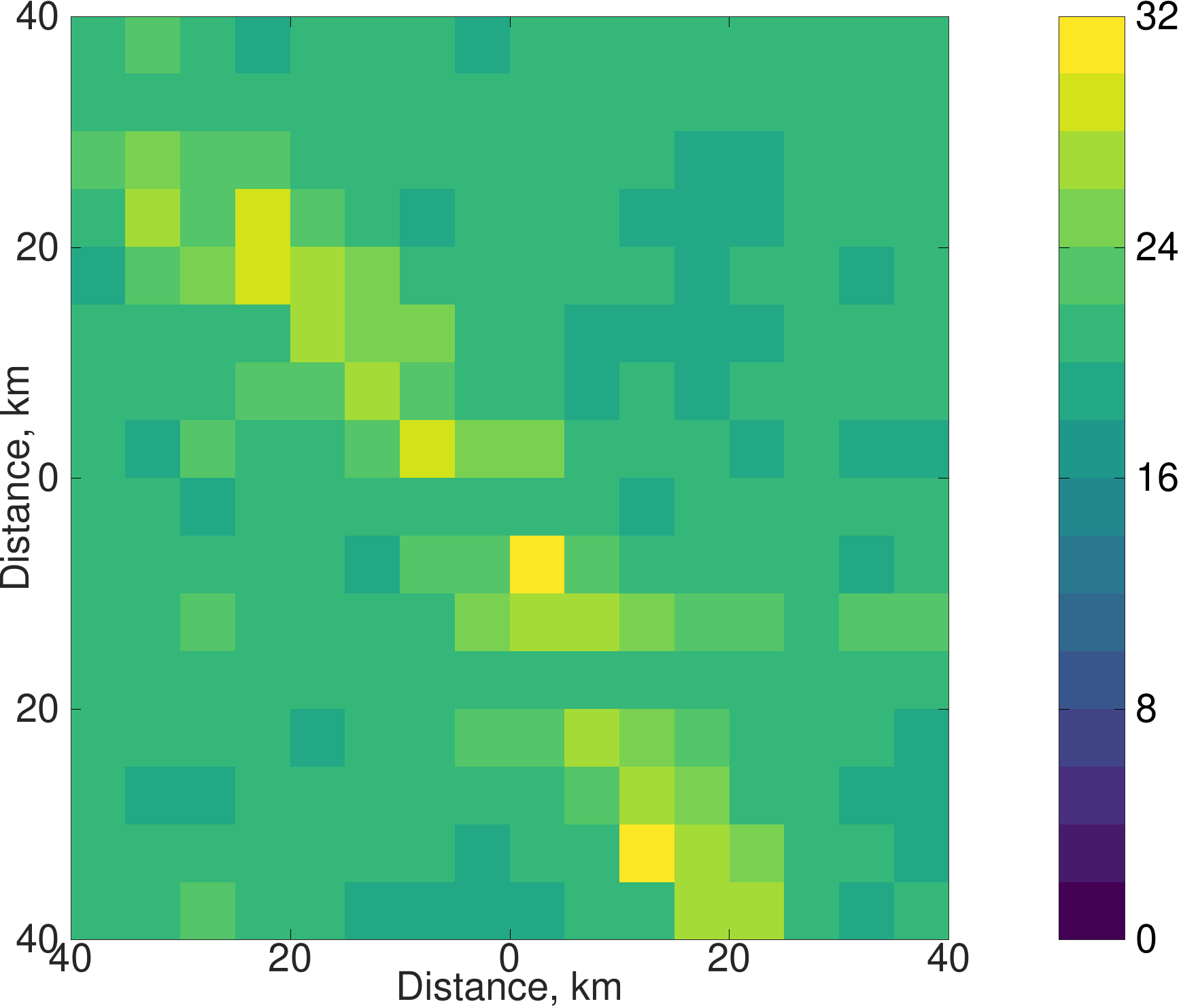}}
	
	\caption{Snapshot of the focal plane for an elve registered on
	September~18, 2016.
	ADC counts are scaled to individual PMT gains.
	Geographic North is approximately at the top of
	the plot. Numbers along the axes denote distance from the center of
	the FOV at sea level. 
	}
	\label{fig:elve}
\end{figure}

TUS has also registered a few events that are likely to be
elves---short-lived optical phenomena that manifest themselves at the
lower edge of the ionosphere as bright rings expanding at the speed of
light up to a maximum radius of $\sim300$~km~\cite{1996GeoRL..23.2157F}.
An event presented in Fig.~\ref{fig:elve} was registered on
September~18, 2016, above Africa ($9^\circ\!\!.7$N, $17^\circ\!\!.1$E)
at 22:06:48~UTC. The record demonstrated an arc-like track that crossed
the focal plane approximately at the speed of light.  Remarkably, the
Vaisala Global Lightning Dataset GLD360~\cite{vaisala1,vaisala2}
registered six lightning discharges in the direction to the center of
the arc within 1~s from the TUS event, with the closest of them in
around $\sim130$~km from the center of the TUS FOV.  The direction to
the lightnings and the geometry and dynamics of the signal support the
elve hypothesis.

\begin{figure}[!ht]
	\cl{\includegraphics[width=0.4\textwidth]{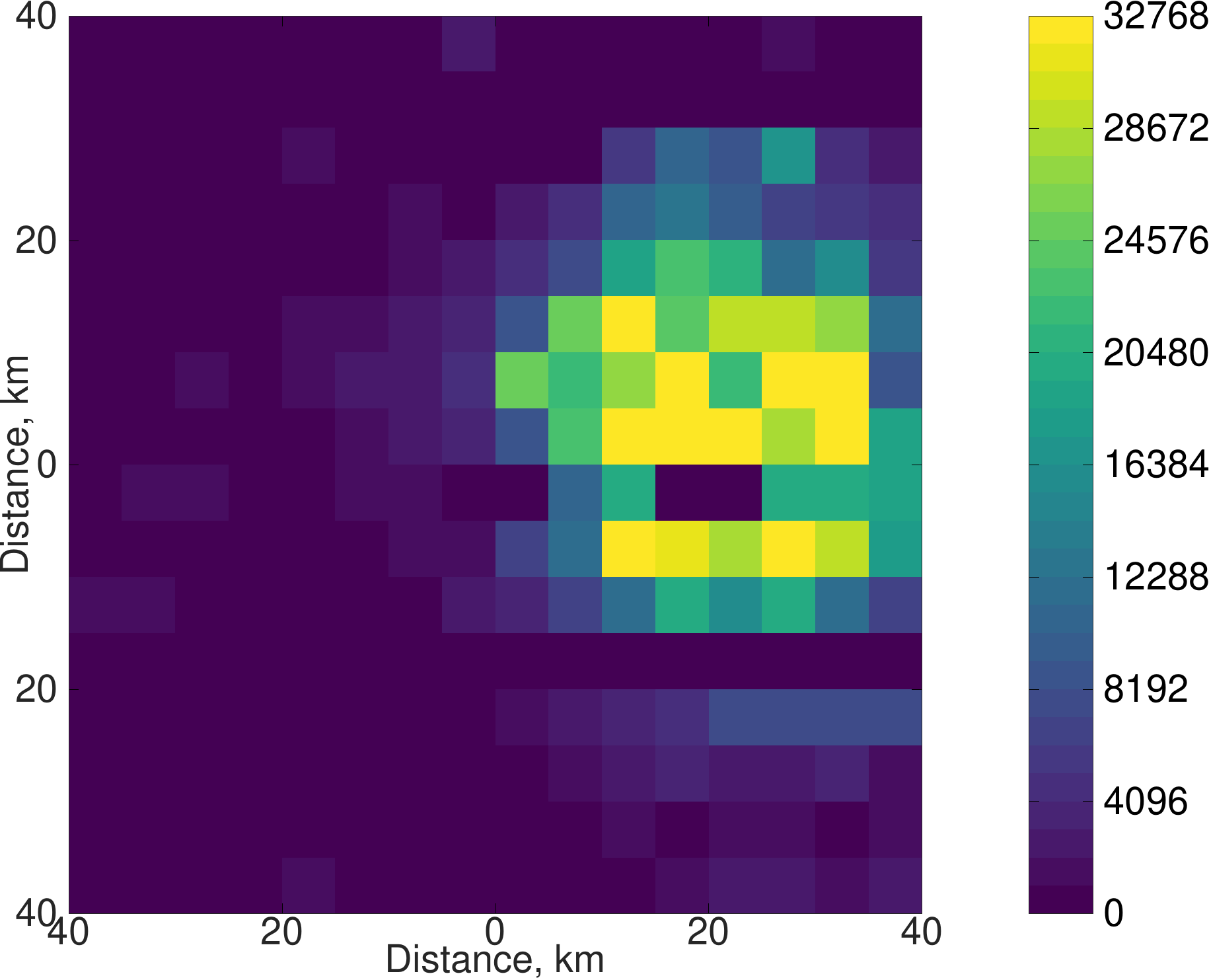}}
	\caption{Snapshot of the focal plane for an event registered
	of June~27, 2016, above India.
	Colours denote real ADC counts.}
	\label{fig:india}
\end{figure}

Numerous thunderstorm events were registered in the TLE mode of
operation with the time sampling of 0.4~ms. These events have diverse
spatial dynamics and temporal structures with multiple peaks and
complicated shapes. Some of them cause signal in all pixels, others
demonstrate a clear structure with non-uniform illumination of the focal
plane and an obvious center of the event. These events were compared
with data of the WWLLN and the Vaisala GLD360.  An example is shown in
Fig.~\ref{fig:india}. It was recorded on June~27, 2016, above India
($25^\circ\!\!.3$N, $77^\circ\!\!.8$E).  Not seen in the figure, the
event had complicated temporal and spatial dynamics.  Several lightning
strikes were registered by the Vaisala GLD360 in this region at the
moment of the event, and two of them took place exactly in the FOV of
TUS.

As we have stated above, the search for an EAS born by an
extreme energy cosmic ray is the priority goal of the TUS experiment.
As of January, 2017, we have not found strong candidates for EAS events
in the TUS data but an intensive study is in progress.

\section{GRB observations on board Lomonosov}

A study of GRBs is one of the main goals of the Lomonosov space mission.
The Lomonosov satellite is the first space mission in which the
multi-wave-length observations of GRBs are realized in real time without
necessity of an optical instrument re-orientation on a GRB monitor
trigger. The mission payload includes the GRB monitor BDRG, wide-field
optical cameras SHOK and the UFFO (Ultra-Fast Flash Observatory)
instrument consisting of the X-ray imaging UBAT telescope and the UV
slewing mirror telescope SMT.  The SHOK cameras are placed in such a way
that their fields of view are overlapped by the GRB monitor detector
FOV.  This allows simultaneous GRB observations in gammas and optics in
all-time scale of event evolution including obtaining optical light
curves of prompt emission as well as of precursors. The real time data
transfer in the Gamma-ray Coordinates Network
(\href{https://gcn.gsfc.nasa.gov/}{GCN}) of detected GRB is realized, as
well as operative control of BDRG data on triggers from ground based
facilities including neutrino and gravitational wave detectors.

The BDRG gamma-ray spectrometer is designed to obtain the temporal and
spectral information of GRBs in the energy range of 10--3000~keV and to
provide GRB triggers on several time scales (10~ms, 1~s and 20~s) for ground and
space telescopes, including UFFO and SHOK. The BDRG instrument consists of
three identical detector boxes with axes shifted by $90^\circ$ from each other.
This configuration allows us to localize a GRB source in the sky with an
accuracy of $\sim2^\circ$. Each BDRG box contains a phoswich NaI(Tl)/CsI(Tl)
scintillator detector.
Data from the three detectors are
collected in an information unit, which generates a GRB trigger and a
set of data frames in the output format. The scientific data output is
$\sim500$~Mb per day, including $\sim180$~Mb of continuous data for events with
duration in excess of 100~ms for 16 channels in each detector, detailed energy
spectra, and sets of frames with $\sim5$~Mb of detailed information for each
burst-like event. 
A number of pre-flight tests including those for the trigger
algorithm and calibration were carried out to confirm the reliability of the
BDRG for operation in space. 

Another instrument consists of two fast, fixed, very
wide-field SHOK cameras. The main goal of this experiment is the GRB optical
emission registration before, synchronously, and after the gamma-ray emission.
The FOV of each camera is placed in the gamma-ray burst
detection area of other devices located on-board Lomonosov.
SHOK provides a registration of optical emissions with a magnitude limit of
9--10$^\mathrm{m}$ on a single frame with an exposure of 0.2~s. The device is
designed for continuous sky monitoring at optical wavelengths in the very wide
field of view (1000 square degrees each camera), detection and localization of
fast time-varying (transient) optical sources on the celestial sphere,
including provisional and synchronous time recording of optical emissions from
the gamma-ray burst error boxes, registered by the BDRG device and implemented
by a control signal (alert trigger) from the BDRG.
The core of each SHOK camera is a
fast-speed 11-Megapixel CCD. Each of the SHOK devices represents a mono-block,
consisting of a node for registration of optical emission, the electronics node,
elements of the mechanical construction, and the body.

The primary aim of UFFO/Lomonosov is to
capture the rise phase of the optical light curve, one of the least
known aspects of GRBs. Fast response measurements of the optical
emission of GRBs are to be made by SMT, which
employs a rapidly slewing mirror to redirect the optical axis of the
telescope to a GRB position prior determined by the UFFO Burst Alert
Telescope (UBAT), the other onboard instrument, for the observation and
imaging of X-rays.

\begin{figure}[!ht]
	\cl{\includegraphics[width=78mm]{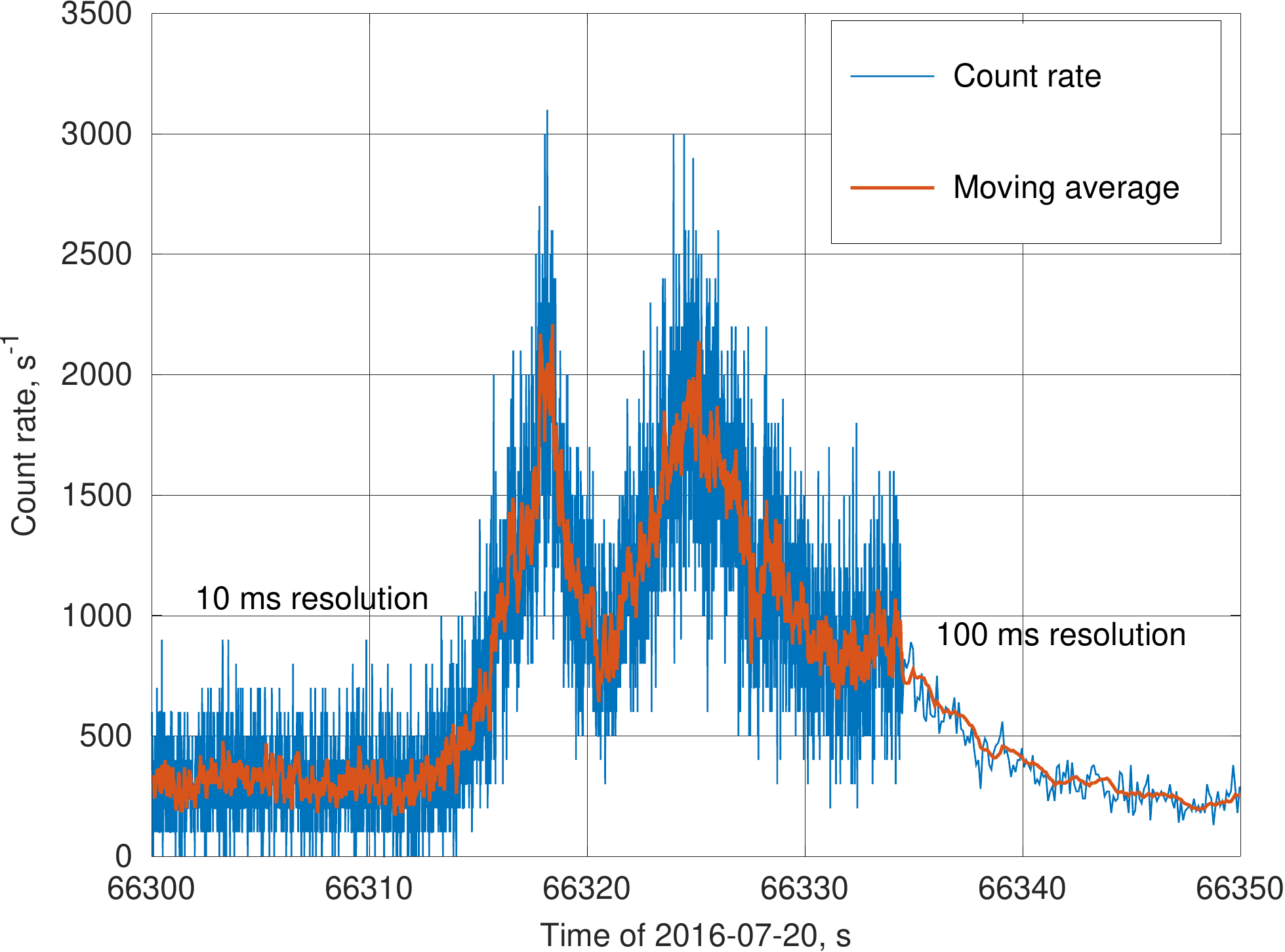}}
	\caption{%
		Time profile of the count rate in the NaI(Tl) 20--170~keV range of the
		BDRG-2 detector for GRB~160720A.}
	\label{fig:GRB}
\end{figure}

The main role of UBAT is to monitor the X-ray sky, rapidly capture any source
illumination and movement, calculate the source location, and then trigger
SMT to observe the early UV/optical
counterpart. To detect X-rays, UBAT utilizes a scintillation detector
composed of a $6\times6$ array of modules of Yttrium Oxyorthosilicate
crystals arranged in an $8\times8$ pixel array, and multi-anode photo
multiplier tubes. It uses the well-known coded aperture mask technique
to estimate a direction vector of a GRB source in its FOV. All functions
are written in the Field Programmable Gate Array firmware to enable fast
triggering and imaging algorithms.

Seventeen GRBs were detected and confirmed by other experiments up to
January, 2017
(see \href{https://downloader.sinp.msu.ru/grb\_catalog/}{https://downloader.sinp.msu.ru/grb\_catalog/}).
GCN circulars were published for all these GRBs. The smaller
than expected number of detected bursts is due to the reduced time for
real observations caused by the satellite testing operations that were
performed during the initial stage of operation. This also explains the
gaps in our data.
The time profile of gamma quantum counts of one of the
GRBs detected by the BDRG instrument is presented in Fig.~\ref{fig:GRB}.
Apart from the 17 observed GRBs, the BDRG instrument has detected six
bursts from the magnetar SGR1935+2154 as well as a few solar
flares at X- and gamma rays.

The main advantages of the BDRG instrument in comparison with the
Konus-Wind experiments~\cite{Konus} are the possibility of gamma by
gamma reading for detected events and the use of phoswich detectors,
which allow effective separation of gamma quanta from electrons. The
possibility of such a distinction provides an opportunity to the GRB
search as a result of the elimination  of electron precipitation-like
events that might otherwise be mistaken for GRBs. In comparison with
other modern GRB experiments, for example the Fermi Gamma-ray Burst
Monitor~\cite{FermiGBM}, the BDRG payload has comparable sensitivity and
temporal resolution, although a slightly smaller effective area. Because
of the possibility of operative GRB data transfer via the
\href{http://www.globalstar.com}{Globalstar} network to the GCN, the
BDRG GRB monitor provides a good input to contemporary GRB observations. 

\begin{acknowledgments}
The TUS experiment team thanks Robert~Holzworth, the head of
the World Wide Lightning Location Network,
and Vaisala Inc.\ company for providing data employed in the present study.
The work was done with partial financial support from the Russian
Foundation for Basic Research grants No.\ 15-02-05498-a and No.\ 16-29-13065.
The Korean work is supported by the National Research Foundation grants
No.\ 2015R1A2A1A01006870 and No.\ 2015R1A2A1A15055344.
\end{acknowledgments}


\begin{thebibliography}{10}
\expandafter\ifx\csname natexlab\endcsname\relax\def\natexlab#1{#1}\fi
\expandafter\ifx\csname bibnamefont\endcsname\relax
  \def\bibnamefont#1{#1}\fi
\expandafter\ifx\csname bibfnamefont\endcsname\relax
  \def\bibfnamefont#1{#1}\fi
\expandafter\ifx\csname citenamefont\endcsname\relax
  \def\citenamefont#1{#1}\fi
\expandafter\ifx\csname url\endcsname\relax
  \def\url#1{\texttt{#1}}\fi
\expandafter\ifx\csname urlprefix\endcsname\relax\def\urlprefix{URL }\fi
\providecommand{\bibinfo}[2]{#2}
\providecommand{\eprint}[2][]{\url{#2}}

\bibitem[{\citenamefont{Sadovnichiy et~al.}(2013)\citenamefont{Sadovnichiy,
  Amelyushkin, Angelopoulos et~al.}}]{Lomonosov2013}
\bibinfo{author}{\bibfnamefont{V.~A.} \bibnamefont{Sadovnichiy}},
  \bibinfo{author}{\bibfnamefont{A.~M.} \bibnamefont{Amelyushkin}},
  \bibinfo{author}{\bibfnamefont{V.}~\bibnamefont{Angelopoulos}},
  \bibnamefont{et~al.}, \bibinfo{journal}{Cosmic Research}
  \textbf{\bibinfo{volume}{51}}, \bibinfo{pages}{427} (\bibinfo{year}{2013}),
  ISSN \bibinfo{issn}{1608-3075}.

\bibitem[{\citenamefont{{Khrenov} et~al.}(2001)\citenamefont{{Khrenov},
  {Panasyuk}, {Alexandrov} et~al.}}]{2001AIPC..566...57K}
\bibinfo{author}{\bibfnamefont{B.~A.} \bibnamefont{{Khrenov}}},
  \bibinfo{author}{\bibfnamefont{M.~I.} \bibnamefont{{Panasyuk}}},
  \bibinfo{author}{\bibfnamefont{V.~V.} \bibnamefont{{Alexandrov}}},
  \bibnamefont{et~al.}, in \emph{\bibinfo{booktitle}{Observing Ultrahigh Energy
  Cosmic Rays from Space and Earth}}, edited by
  \bibinfo{editor}{\bibfnamefont{H.}~\bibnamefont{{Salazar}}},
  \bibinfo{editor}{\bibfnamefont{L.}~\bibnamefont{{Villasenor}}},
  \bibnamefont{and} \bibinfo{editor}{\bibfnamefont{A.}~\bibnamefont{{Zepeda}}}
  (\bibinfo{year}{2001}), vol. \bibinfo{volume}{566} of
  \emph{\bibinfo{series}{American Institute of Physics Conference Series}}, pp.
  \bibinfo{pages}{57--75}.

\bibitem[{\citenamefont{{Khrenov} et~al.}(2013)\citenamefont{{Khrenov},
  {Panasyuk}, {Garipov} et~al.}}]{TUS-ecrs2012}
\bibinfo{author}{\bibfnamefont{B.~A.} \bibnamefont{{Khrenov}}},
  \bibinfo{author}{\bibfnamefont{M.~I.} \bibnamefont{{Panasyuk}}},
  \bibinfo{author}{\bibfnamefont{G.~K.} \bibnamefont{{Garipov}}},
  \bibnamefont{et~al.}, in \emph{\bibinfo{booktitle}{European Physical Journal
  Web of Conferences}} (\bibinfo{year}{2013}), vol.~\bibinfo{volume}{53}, p.
  \bibinfo{pages}{09006}.

\bibitem[{\citenamefont{{Garipov} et~al.}(2017)\citenamefont{{Garipov},
  {Zotov}, {Klimov} et~al.}}]{tracks-izvran}
\bibinfo{author}{\bibfnamefont{G.~K.} \bibnamefont{{Garipov}}},
  \bibinfo{author}{\bibfnamefont{M.~Y.} \bibnamefont{{Zotov}}},
  \bibinfo{author}{\bibfnamefont{P.~A.} \bibnamefont{{Klimov}}},
  \bibnamefont{et~al.}, \bibinfo{journal}{Bull. Rus. Acad. Sci. Physics}
		\textbf{\bibinfo{volume}{81}}, \bibinfo{pages}{407}
		(\bibinfo{year}{2017}).

\bibitem[{\citenamefont{{Agostinelli} et~al.}(2003)\citenamefont{{Agostinelli},
  {Allison}, {Amako} et~al.}}]{Geant4}
\bibinfo{author}{\bibfnamefont{S.}~\bibnamefont{{Agostinelli}}},
  \bibinfo{author}{\bibfnamefont{J.}~\bibnamefont{{Allison}}},
  \bibinfo{author}{\bibfnamefont{K.}~\bibnamefont{{Amako}}},
  \bibnamefont{et~al.}, \bibinfo{journal}{Nuclear Instruments and Methods in
  Physics Research A} \textbf{\bibinfo{volume}{506}}, \bibinfo{pages}{250}
  (\bibinfo{year}{2003}).

\bibitem[{\citenamefont{{Fukunishi} et~al.}(1996)\citenamefont{{Fukunishi},
  {Takahashi}, {Kubota} et~al.}}]{1996GeoRL..23.2157F}
\bibinfo{author}{\bibfnamefont{H.}~\bibnamefont{{Fukunishi}}},
  \bibinfo{author}{\bibfnamefont{Y.}~\bibnamefont{{Takahashi}}},
  \bibinfo{author}{\bibfnamefont{M.}~\bibnamefont{{Kubota}}},
  \bibnamefont{et~al.}, \bibinfo{journal}{Geophysical Research Letters}
  \textbf{\bibinfo{volume}{23}}, \bibinfo{pages}{2157} (\bibinfo{year}{1996}).

\bibitem[{\citenamefont{{Said} et~al.}(2010)\citenamefont{{Said}, {Inan}, and
  {Cummins}}}]{vaisala1}
\bibinfo{author}{\bibfnamefont{R.~K.} \bibnamefont{{Said}}},
  \bibinfo{author}{\bibfnamefont{U.~S.} \bibnamefont{{Inan}}},
  \bibnamefont{and} \bibinfo{author}{\bibfnamefont{K.~L.}
  \bibnamefont{{Cummins}}}, \bibinfo{journal}{Journal of Geophysical Research
  (Atmospheres)} \textbf{\bibinfo{volume}{115}}, \bibinfo{eid}{D23108}
  (\bibinfo{year}{2010}).

\bibitem[{\citenamefont{{Said} et~al.}(2013)\citenamefont{{Said}, {Cohen}, and
  {Inan}}}]{vaisala2}
\bibinfo{author}{\bibfnamefont{R.~K.} \bibnamefont{{Said}}},
  \bibinfo{author}{\bibfnamefont{M.~B.} \bibnamefont{{Cohen}}},
  \bibnamefont{and} \bibinfo{author}{\bibfnamefont{U.~S.}
  \bibnamefont{{Inan}}}, \bibinfo{journal}{Journal of Geophysical Research
  (Atmospheres)} \textbf{\bibinfo{volume}{118}}, \bibinfo{pages}{6905}
  (\bibinfo{year}{2013}).

\bibitem[{\citenamefont{{Mazets} et~al.}(2012)\citenamefont{{Mazets},
  {Aptekar}, {Golenetskii} et~al.}}]{Konus}
\bibinfo{author}{\bibfnamefont{E.~P.} \bibnamefont{{Mazets}}},
  \bibinfo{author}{\bibfnamefont{R.~L.} \bibnamefont{{Aptekar}}},
  \bibinfo{author}{\bibfnamefont{S.~V.} \bibnamefont{{Golenetskii}}},
  \bibnamefont{et~al.}, \bibinfo{journal}{Sov. J. Exp.
  Theor. Phys. Letters} \textbf{\bibinfo{volume}{96}},
  \bibinfo{pages}{544} (\bibinfo{year}{2012}).

\bibitem[{\citenamefont{{Meegan} et~al.}(2009)\citenamefont{{Meegan}, {Lichti},
  {Bhat} et~al.}}]{FermiGBM}
\bibinfo{author}{\bibfnamefont{C.}~\bibnamefont{{Meegan}}},
  \bibinfo{author}{\bibfnamefont{G.}~\bibnamefont{{Lichti}}},
  \bibinfo{author}{\bibfnamefont{P.~N.} \bibnamefont{{Bhat}}},
  \bibnamefont{et~al.}, \bibinfo{journal}{\apj} \textbf{\bibinfo{volume}{702}},
  \bibinfo{eid}{791-804} (\bibinfo{year}{2009}), \eprint{0908.0450}.

\end{thebibliography}
\end{document}